\documentclass[onecolumn,preprint,preprintnumbers,amsmath,amssymb,prb,apsrev4-2]{revtex4-2}
\usepackage{graphicx}
\usepackage{dcolumn}
\usepackage{bm}
\usepackage{SIunits}
\usepackage{verbatim}
\usepackage{placeins}
\usepackage{natbib}
\usepackage{amsmath}
\usepackage{amssymb}
\usepackage{longtable}
\usepackage{tabularx}
\usepackage{keyval}
\usepackage{multirow}
\usepackage[colorlinks,linkcolor=blue]{hyperref}

\begin{document}
\title{Chirality-selective topological magnon phase transition induced by interplay of anisotropic exchange interactions in honeycomb ferromagnet}
\author{Jin-Yu Ni$^{1,2}$}
\author{Xia-Ming Zheng$^{1}$}
\author{Peng-Tao Wei$^{1}$}
\author{Da-Yong Liu$^{3,1}$}
\email[]{dyliu@ntu.edu.cn}
\author{Liang-Jian Zou$^{1}$}
\email[]{zou@theory.issp.ac.cn}
\affiliation{1 Key Laboratory of Materials Physics, Institute of Solid State Physics, HFIPS, Chinese Academy of Sciences, Hefei 230031, China}
\affiliation{2 Science Island Branch of Graduate School, University of Science and Technology of China, Hefei 230026, China}
\affiliation{3 Department of Physics, School of Physics and Technology, Nantong University, Nantong 226019, China} 
\href{http://orcid.org/0000-0003-4370-473X}{ORCiD: 0000-0003-4370-473X}

\date{\today}

\begin{abstract} 
A variety of distinct anisotropic exchange interactions commonly exist in one magnetic material due to complex crystal, magnetic and orbital symmetries. Here we investigate the effects of multiple anisotropic exchange interactions on topological magnon in a honeycomb ferromagnet, and find a chirality-selective topological magnon phase transition induced by a complicated interplay of Dzyaloshinsky-Moriya interaction (DMI) and pseudo-dipolar interaction (PDI), accompanied by the bulk gap close and reopen with chiral inversion.
Moreover, this novel topological phase transition involves band inversion at high symmetry points $K$ and $K'$, which can be regarded as a pseudo-orbital reversal, {\it i.e.} magnon valley degree of freedom, implying a new manipulation corresponding to a sign change of the magnon thermal Hall conductivity. Indeed, it can be realized in 4$d$ or 5$d$ correlated materials with both spin-orbit coupling and orbital localized states, such as iridates and ruthenates, {\it etc}. This novel regulation may have potential applications on magnon devices and topological magnonics.
\end{abstract}

\vskip 300 pt

\maketitle

\section{Introduction}
Topological matters have been a hot topic since the discovery of topological insulators \cite{RMP82-3045,RMP83-1057,Spin01-33}. With the development of topological band theory, rich topological states have been proposed. Since the topological band structure is independent of the statistics of the constituent particles, besides topological fermions of electronic origin, topological bosons also attract much attention, such as phonons \cite{Nat555-342,PRL103-248101}, photons \cite{NM12-233,RMP91-015006,Sci359-eaar4005}, magnons \cite{PRB87-144101,JPCM28-386001,NC9-2591,JPCM29-385801,PRA9-024029,JAP129-151101,PR915-1} {\it etc}. Particularly, it is known that the magnon is the quantum of the spin wave, that is the spin excitation of a magnetically ordered system. Analog to the topological electronic states, various topological magnons, {\it e.g.} topological magnon insulators, Dirac and Weyl magnons \cite{PRL119-247202,PRX8-011010,NC7-12691,PRB94-075401,PRB117-157204}, have been introduced in different types of lattices theoretically and experimentally, such as honeycomb \cite{JPCM28-386001,PRB95-014435,PELDSN135-114984,PRL127-217202}, Kagome \cite{PRB87-144101,PRL115-147201,PRB90-024412,PRB104-144422}, triangular \cite{PRB100-064412,JPCM29-385801}, and Lieb lattices \cite{JPCM27-166003}. Moreover, the experimentally observed magnon thermal Hall effect in the insulated ferromagnet Lu$_{2}$V$_{2}$O$_{7}$ with pyrochlore lattice \cite{Sci329-297} greatly promotes the potential applications of magnon in devices, {\it i.e.} magnonics \cite{JPD43-264001,NP11-453}. Matsumoto and Murakami \cite{PRL106-197202,PRB84-184406} explained this phenomenon as a result of non-compensated magnon edge currents by the semiclassical picture. Later Zhang {\it et al.} \cite{PRB87-144101} attributed it to the topology nature.

Different from the electronic system, magnons originating from more magnetic structures than crystal structures can be tunably manipulated by distinct magnetic interactions, for instance, isotropic Heisenberg exchange interaction, axial anisotropic interaction, external magnetic field, and anisotropic interactions including dipole-dipole interaction, Kitaev interaction \cite{AP321-2}, Dzyaloshinskii-Moriya interaction (DMI) \cite{JPCS4-241,PR120-91}, and pseudo-dipolar interaction (PDI) \cite{PR52-1178,PRL102-017205}, {\it etc}. Notice that in electronic systems, spin-orbit coupling plays an important role in topological electronic states, such as opening a gap in topological insulators. Analogously, anisotropic interactions as a magnonic version of spin-orbit coupling play an essential role in topological magnons. As is well known, the next nearest-neighboring (NNN) DMI from the spin-orbit coupling due to the inversion symmetry broken on honeycomb lattice often leads to a topological magnon insulator with topologically protected edge states characterized by nonzero Chern number. In addition, the nearest-neighboring (NN) PDI \cite{PR52-1178} from the combined effects of the spin-orbit coupling and orbital quenching on honeycomb lattice has been found also introducing topology on magnons \cite{PRB95-014435}.

In general, two or more anisotropic interactions generally exist in realistic materials due to complex lattices and magnetic symmetries.
For instance, the recently discovered abundant 4$d$ and 5$d$ correlated materials, such as iridates and ruthenates, most possibly exhibit multiple anisotropic interactions due to the coexistence of the spin-orbit coupling and orbital interaction \cite{PRL108-127204}. 
It has been demonstrated that either DMI or PDI can separately lead to topological magnonic states. However, it is unclear whether the combination of them will still lead to topological magnonic states since their interplay may be competitive or cooperative due to their different origins. Therefore, the general phase diagram dependent on the multiple anisotropic interactions is worth investigating.

Here we mainly investigate the effect of the combined anisotropic DMI and PDI on the topological magnons in honeycomb ferromagnet, and find a topological phase diagram with a sign change of Chern number of the conduction band attributing from the competitive anisotropic interactions of DMI and PDI. Meanwhile, the related topological properties, {\it e.g.} Chern number, Berry curvature, magnon thermal Hall conductivity, spin structure factor, and edge states of the topological phases, are presented. The rest of this paper is organized as follows: first, the model and method are introduced in Sec. II; then the result and discussion are given in Sec. III; and the final section is devoted to the conclusion.

\section{Model and Method}
\begin{figure}[htbp]
\hspace*{-2mm}
\centering
\includegraphics[trim = 0mm 0mm 0mm 0mm, clip=true, angle=0, width=1.0 \columnwidth]{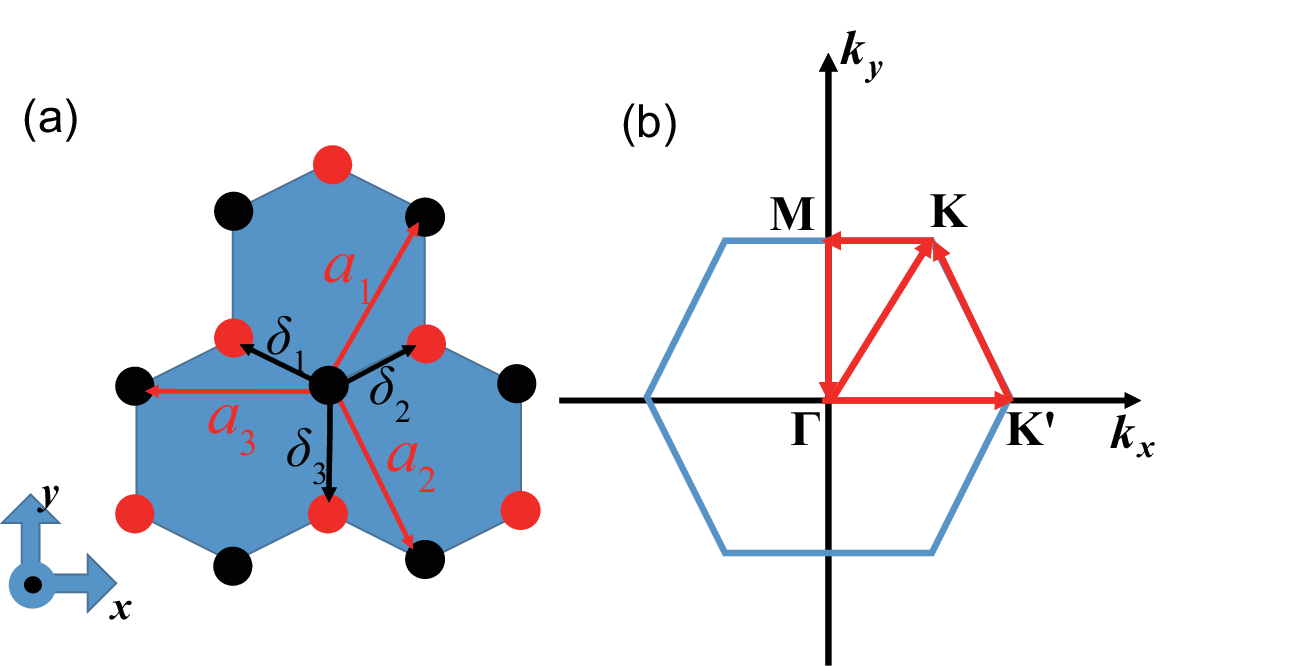}
\caption{(Color online) (a) The honeycomb lattice with the NN vector $\mathbf{\delta_{1,2,3}}$ and the NNN vector $\mathbf{a_{1,2,3}}$ in the $x$-$y$ plane. Red and black dots indicate the sublattices A and B, respectively. (b) The corresponding Brillouin zone (BZ), and its high-symmetry $k$-points and $k$-paths.}
\label{Fig1}
\end{figure}

We consider a two-dimensional (2D) Heisenberg ferromagnet on a honeycomb lattice, as displayed in Fig.~\ref{Fig1}, including both the isotropic and the anisotropic exchange interactions.
The model Hamiltonian is described as \cite{JPCM28-386001,PRL125-217202}
\begin{equation}\label{1}
    \begin{split}
  H=&-J\sum\limits_{\langle i,j\rangle}\mathbf{S_{i}}\cdot \mathbf{S_{j}}-F\sum\limits_{\langle i,j\rangle}(\mathbf{S_{i}}\cdot \mathbf{e_{ij}})(\mathbf{S_{j}}\cdot \mathbf{e_{ij}})\\\\
  &+\sum\limits_{\langle\langle i,j\rangle\rangle}\mathbf{D_{ij}}\cdot (\mathbf{S_{i}}\times \mathbf{S_{j}})-A_{z}\sum\limits_{i}(\mathbf{S_{i}^{z}})^{2}\\\\
  &-g\mu_{B}B_{z}\sum\limits_{i}\mathbf{S_{i}^{z}}
    \end{split}
\end{equation}
where $\langle i,j\rangle$ denotes the NN sites, $\langle\langle i,j\rangle\rangle$ the NNN ones. The first two terms are the Heisenberg-type and pseudo-dipolar exchange interactions with exchange constants $J$ and $F$, respectively. The unit vector $\mathbf{e_{ij}}$ connects sites $i$ and $j$ chosen from three NN vectors $\mathbf{\delta_{1,2,3}}$. The third term is the NNN ($\mathbf{a_{1,2,3}}$) DMI arising from the lack of inversion symmetry and only the out-of-plane direction ($z$) component is taken into account, $\mathbf{D_{ij}}=v_{ij}D_{z}\mathbf{\widehat{z}}$, where $\mathbf{\widehat{z}}$ is the unit vector along the $z$ direction, and $v_{ij}$=+1 for sublattice A and $-$1 for sublattice B. Obviously, the DMI breaks the chiral symmetry. The last two terms are the single-ion anisotropy (SIA) with the easy-axis along the $z$ direction, and the Zeeman term from magnetic field $B$ along the $z$ direction, respectively. 
Because the SIA $A_{z}$ term and the external magnetic field $B_{z}$ term only shift the total energy level, we find when $2A_{z}+g\mu_{B}B_{z}/S>3F/2$ the ferromagnetic ground state can be stabilized through estimating the energy at $\Gamma$ point. Notice that in realistic systems, the values of single ion anisotropic parameters $A_{z}$ may vary with materials and external fields. For convenience, we fix $A_{z}$=10$J$ and $B_{z}$=0 (absence of external magnetic field) in all the calculations to stabilize the ferromagnetic ground state.

Note that in realistic one-dimensional (1D) and 2D magnetic materials, there exist very strong anisotropic magnetic interactions due to the anisotropic origin from the strong spin-orbit coupling and orbital ordering, indicating that very large ratios of $A_{z}$, $D_{z}$ and $F$ to $J$ can be achieved. Indeed, there are already first-principle calculations and experimental evaluations of strong SIA, DMI and PDI \cite{PRL125-217202,NPJCM6-158}. In particular, it is experimentally discovered that the PDI parameter $F$ can be about 16, 25, and 30 times larger than the Heisenberg exchange $J$ in $\alpha$-RuCl$_{3}$ \cite{NC11-1603}, CrI$_{3}$ \cite{PRL124-017201}, Na$_{2}$IrO$_{3}$ \cite{PRL105-027204,JPCM29-493002,PRX9-031047}, respectively. Thus the strength of PDI ($|F/J|$) can be the strongest up to 30. In addition, large SIA values ($A_{z}$) are also reported in a variety of 2D magnetic materials \cite{PRL125-217202,AEM9-2200650,NPJCM6-158}. We emphasize that the conclusions do not change regardless of the choice of parameters in the following discussion.

To obtain the magnon spectrum, the standard linear Holstein-Primakoff (HP) transformation \cite{PR58-1098} is performed.
\begin{equation}\label{2}
S_{i}^z=S-a_{i}^{\dag} a_{i},S_{i}^{+}=\sqrt{2S}a_{i},S_{i}^{-}=\sqrt{2S}a_{i}^{\dag}
\end{equation}
where $a_{i}^{\dag}$ and $a_{i}$ ($b_{i}^{\dag}$ and $b_{i}$) are magnon creation and annihilation operators in the A (B) sublattice, and $S_{i}^{\pm}=S_{i}^{x}\pm iS_{i}^{y}$ denote the spin raising and lowering operators. After carrying out a Fourier transformation into momentum space, a 4$\times$4 Hamiltonian matrix $H=const.+\sum_{\mathbf{k}} \Psi_{\mathbf{k}}^{\dag} H(\mathbf{k}) \Psi_{\mathbf{k}}$ with $\Psi_{\mathbf{k}}^{\dag}=(a_{\mathbf{k}}^{\dag},a_{-\mathbf{k}},b_{\mathbf{k}}^{\dag},b_{-\mathbf{k}})$ is obtained. The Hamiltonian matrix $H(\mathbf{k})$ is written as
\begin{equation}
H=\left(
    \begin{array}{cccc}
    h_{A}(\mathbf{k}) & 0 & -f(\mathbf{k}) & g_{1}^{\ast}(\mathbf{k})\\
    0        & h_{B}(\mathbf{k}) & g_{2}(\mathbf{k}) & -f(\mathbf{k})\\
    -f^{\ast}(\mathbf{k}) & g_{2}^{\ast}(\mathbf{k}) & h_{B}(\mathbf{k}) & 0\\
    g_{1}(\mathbf{k}) & -f^{\ast}(\mathbf{k}) & 0 & h_{A}(\mathbf{k})
    \end{array}\right)
\end{equation}
where $h_{A}(\mathbf{k})=3JS+2A_{z}S+g\mu_{B}B_{z}-2D_{z}S\sum\limits_{i}sin(\mathbf{k\cdot a_{i}})$, $h_{B}(\mathbf{k})=3JS+2A_{z}S+g\mu_{B}B_{z}+2D_{z}S\sum\limits_{i}sin(\mathbf{k\cdot a_{i}})$, $f(\mathbf{k})=(J+\frac{F}{2})S\sum\limits_{j}e^{i\mathbf{k\cdot \delta_{j}}}$, $g_{1}(\mathbf{k})=-\frac{FS}{2}\sum\limits_{j}e^{2i\theta_{j}}e^{i\mathbf{k\cdot \delta_{j}}}$, $g_{2}(\mathbf{k})=-\frac{FS}{2}\sum\limits_{j}e^{-2i\theta_{j}}e^{i\mathbf{k\cdot \delta_{j}}}$, $\theta_{j}$ is the angle between $\mathbf{\delta_{j}}$ and the $x$ direction. So we can perform diagonalization and obtain the eigenvalues and eigenvectors. Note that all magnetic coupling parameters ($D_{z}$ and $F$) are in unit of $J$.

\section{Result and Discussion}
\subsection{\it Competitive phase diagram}
\begin{figure}[htbp]
\hspace*{-2mm}
\centering
\includegraphics[trim = 0mm 0mm 0mm 0mm, clip=true, angle=0, width=1.0 \columnwidth]{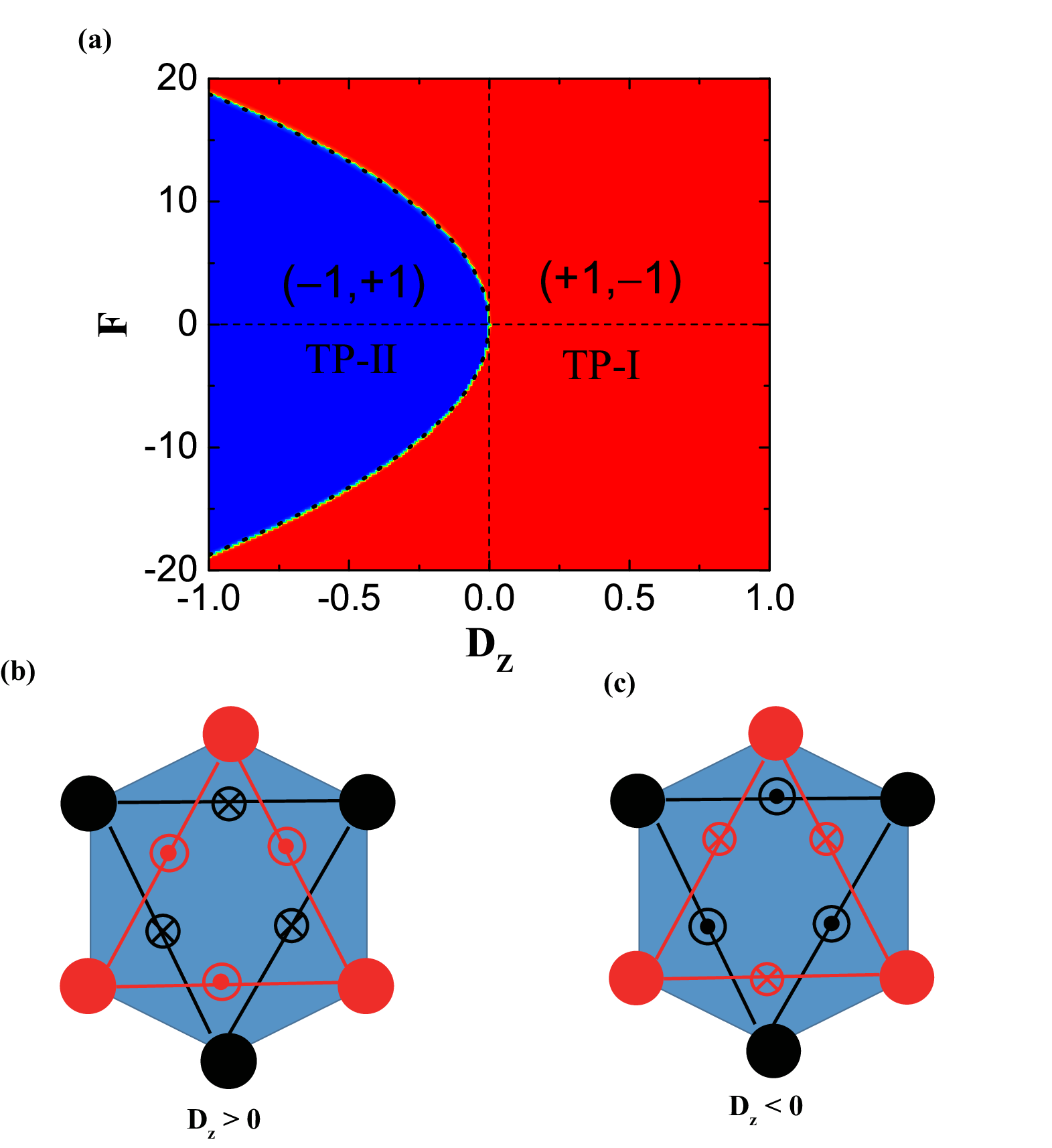}
\caption{(Color online) (a)$D_{z}$-$F$ phase diagram for magnons in honeycomb ferromagnet. The symbols ($+/-$1, $-/+$1) denote the Chern number of conduction and valence bands, respectively. The dot line on the boundary is our analytical result. The directions of DMI for $D_{z}$ $>$ 0 (b) and $D_{z}$ $<$ 0 (c). The crossed and dotted circles denote the alternating DMI along the NNN bonds.}
\label{Fig2}
\end{figure}

In this paper we focus on the interplay of anisotropic DMI and PDI, since the sole DMI and PDI can individually induce a magnon topological insulating phase on the 2D honeycomb lattice, which is distinctly different from that on the 1D Su–Schrieffer–Heeger (SSH) lattice \cite{JPCM34-495801}. The $D_{z}$-$F$ phase diagram for the topological phase of magnons is displayed in Fig.~\ref{Fig2}. Competitive magnon topological phases are found dependent on the sign of DMI $D_{z}$ with the chiral characteristic, indicating a chirality-selective. For positive $D_{z}$, it is a topological phase (TP-I phase) with $+$1 Chern number of the conduction band, and there is no phase transition regardless of the value $F$ of PDI. While for negative $D_{z}$, a new topological phase (TP-II phase) with $-$1 Chern number of conduction band emerges for relatively weak PDI. Obviously, a topological phase transition occurs induced by the complicated interplay of DMI and PDI. Therefore, it is demonstrated that the interplay of DMI and PDI behaves as not simply reinforcing or canceling, but either competing or cooperating, depending on the chiral characteristic of the DMI, {\it i.e.} the sign of the $D_{z}$. Such chirality-dependent topological magnon phase transition can be easily understood based on the band structure character, as discussed in the following.

\subsection{\it Magnon spectrum}
\begin{figure}[htbp]
\hspace*{-2mm}
\centering\includegraphics[trim = 0mm 0mm 0mm 0mm, clip=true, angle=0, width=0.8 \columnwidth]{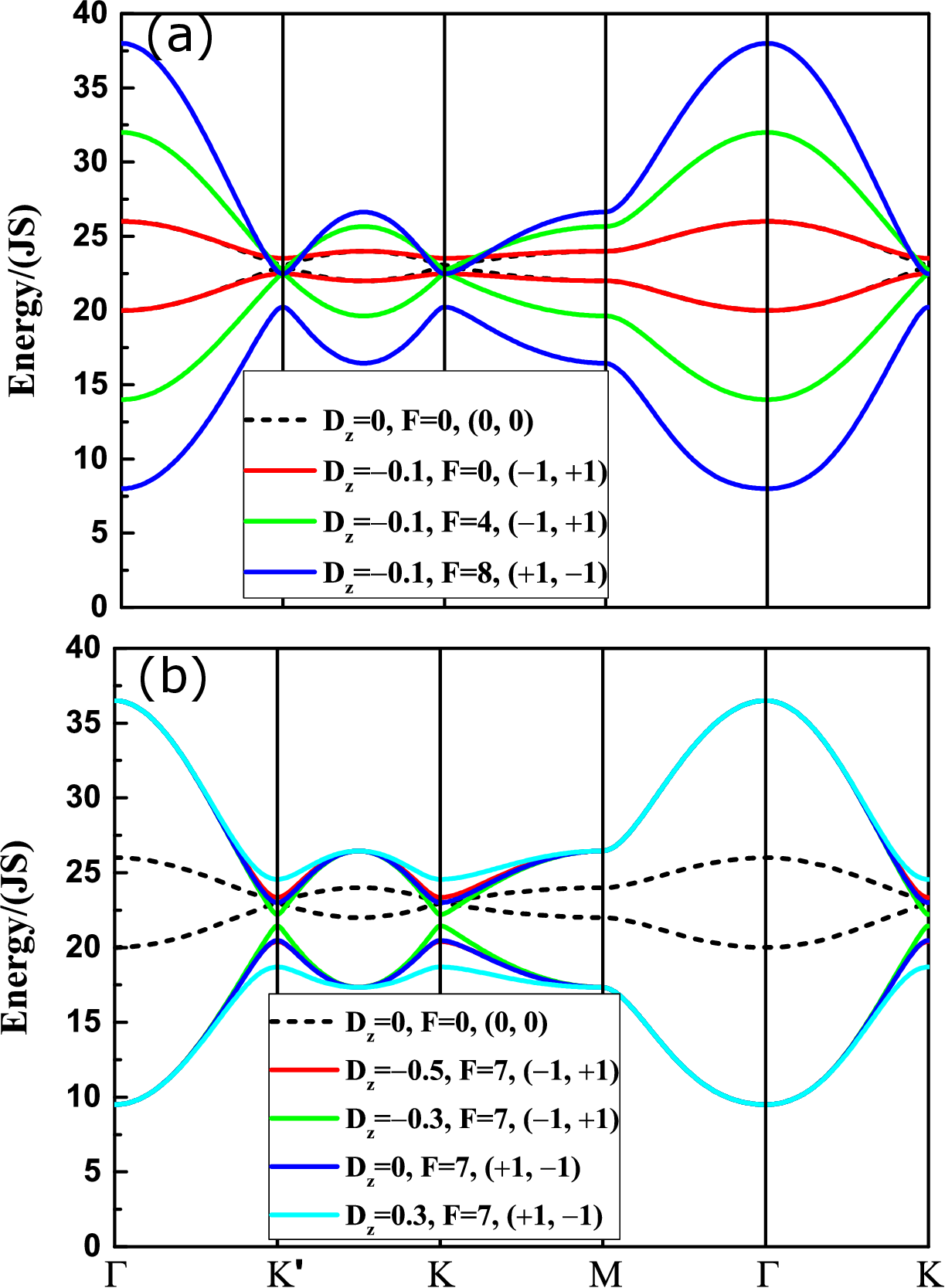}
\caption{(Color online) The evolution of magnon band structures depends on different DMI and PDI parameters: (a) $D_{z}$=$-$0.1 is fixed, and $F$=0, 4, 8, respectively; (b) $F$=7 is fixed, and $D_{z}$=$-$0.5, $-$0.3, 0, 0.3, respectively. The symbols ($+/-$1, $-/+$1) denote the Chern number of conduction and valence bands, respectively. The dash line represents the $D_{z}$=$F$=0 case.}
\label{Fig3}
\end{figure}

To explore the details of the topological phase transition, the evolution of the magnon spectrum on anisotropic DMI $D_{z}$ and PDI $F$ is presented in Fig.~\ref{Fig3} (a) and (b). It can be clearly found that there are Dirac dispersions at high symmetry $K$ and $K^{'}$ points in the absence of both DMI and PDI. Although both DMI and PDI can open a gap at $K$ and $K^{'}$ points, inducing a topological magnon insulating state, they show distinct behaviors on band dispersion. The chiral DMI just only affects the dispersions around $K$ and $K^{'}$ points, while the PDI has great influence on the whole Brillouin zone (BZ) dispersions. Especially, pure DMI can induce flat bands around $K$, $K^{'}$ and $M$ but not $\Gamma$, while PDI enhances the dispersions of the whole bands, which implies a possible competing effect between them. In addition, in accordance with the topological phase transition, the energy gap undergoes a process of closing and reopening, implying a competitive behavior of the anisotropic DMI and PDI. Indeed, the competition of DMI ($D_{z}<0$) and PDI ($F>0$) originates from their opposite effect on the band structure in the momentum $k$-space. All these results also demonstrate that the different anisotropic interactions can be synergistic or competitive, suggesting a new type of manipulation.

On the other hand, through analyzing the evolution of the energy gap at high-symmetry $\Gamma$=(0, 0), $K$=($\frac{2\pi}{3\sqrt{3}}$, $\frac{2\pi}{3}$) and $K^{'}$=($\frac{4\pi}{3\sqrt{3}}$, 0) points, the phase boundary can be determined analytically. For example, the Hamiltonian at $K^{'}$ point can be written as
\begin{equation}
H=\left(
    \begin{array}{cccc}
    A+3\sqrt{3}D_{z}S & 0 & 0 & 0\\
    0 & A-3\sqrt{3}D_{z}S & -\frac{3}{2}FS & 0\\
    0 & -\frac{3}{2}FS & A-3\sqrt{3}D_{z}S & 0\\
    0 & 0 & 0 & A+3\sqrt{3}D_{z}S
    \end{array}\right)
\end{equation}
where $A=3JS+2A_{z}S+g\mu_{B}B_{z}$, so we can perform diagonalization and obtain the eigenvalues and eigenvectors. Then we get the gap at $K^{'}$ point is $\Delta_{g}=A+3\sqrt{3}D_{z}S-\sqrt{(A-3\sqrt{3}D_{z}S)^{2}-(\frac{3}{2}FS)^{2}}$. When $\Delta_{g}=0$, then $\frac{D_{z}}{F^{2}}=-\frac{\sqrt{3}}{16(3J+2A_{z}+g\mu_{B}B_{z}/S)}$. This is the critical phase boundary corresponding to the topological phase transition in Fig.~\ref{Fig2}, since the phase transition is characterized by the gap closing and gap reopening. The same result can be obtained at $K$ point. From the formula of the energy gap, we can see that when $D_{z}>0$ the interplay of the anisotropic DMI and PDI is always synchronous, and when $D_{z}<0$ there is a competition between them. In the same way, the gap at $\Gamma$ point is determined by $\Delta_{g}=6S(J+\frac{F}{2})$, in which the effect of the DMI is absent. Note that this gap closing at $\Gamma$ is related to a trivial state rather than a non-trivial one at $K$ and $K^{'}$, which is verified by our Berry curvature calculation in the following.
Indeed, the interplay between DMI and PDI on the topological magnon phase transition can be exactly analyzed by the zero energy gap condition at high symmetry point $K^{'}$ or $K$ with $\frac{D_{z}}{F^{2}}=-\frac{\sqrt{3}}{16(3J+2A_{z})}$ in the absence of magnetic field $B_{z}$ as mentioned above. It is obviously found that for $D_{z}>0$, the equation can't be realized in the realistic parameter range, indicating the absence of the topological phase transition. While only $D_{z}<0$ corresponds to one chiral symmetry, the topological phase transition emerges, implying a competition between DMI and PDI. Here we emphasize that the occurrence of topological phase transitions does not necessarily depend on large DMI or PDI parameters, but on the ratio of $\frac{D_{z}}{F^{2}}$. In other words, for the realistic materials with small $D_{z}$ values, $F$ values will also be very small. It is clearly demonstrated that the interplay of these two anisotropic magnetic interactions behaves as either cooperative or competitive which depends on the chiral DMI, and naturally induces a chirality-selective topological magnon phase transition. For comparison, the 1D SSH lattice \cite{JPCM34-495801} possesses completely distinct topological states in the presence of both DMI and PDI. This is mainly because the non-trivial energy gap is only associated with the local high-symmetry k points ({\it e.g.} $K$ and $K^{'}$ in honeycomb lattice, $k$=$\pi$ ($-\pi$) in SSH lattice due to different lattice symmetries) in momentum space.

\subsection{\it Pseudo-orbital reversal}
\begin{figure}[htbp]
\hspace*{-2mm}
\centering
\includegraphics[trim = 0mm 0mm 0mm 0mm, clip=true, angle=0, width=1.0 \columnwidth]{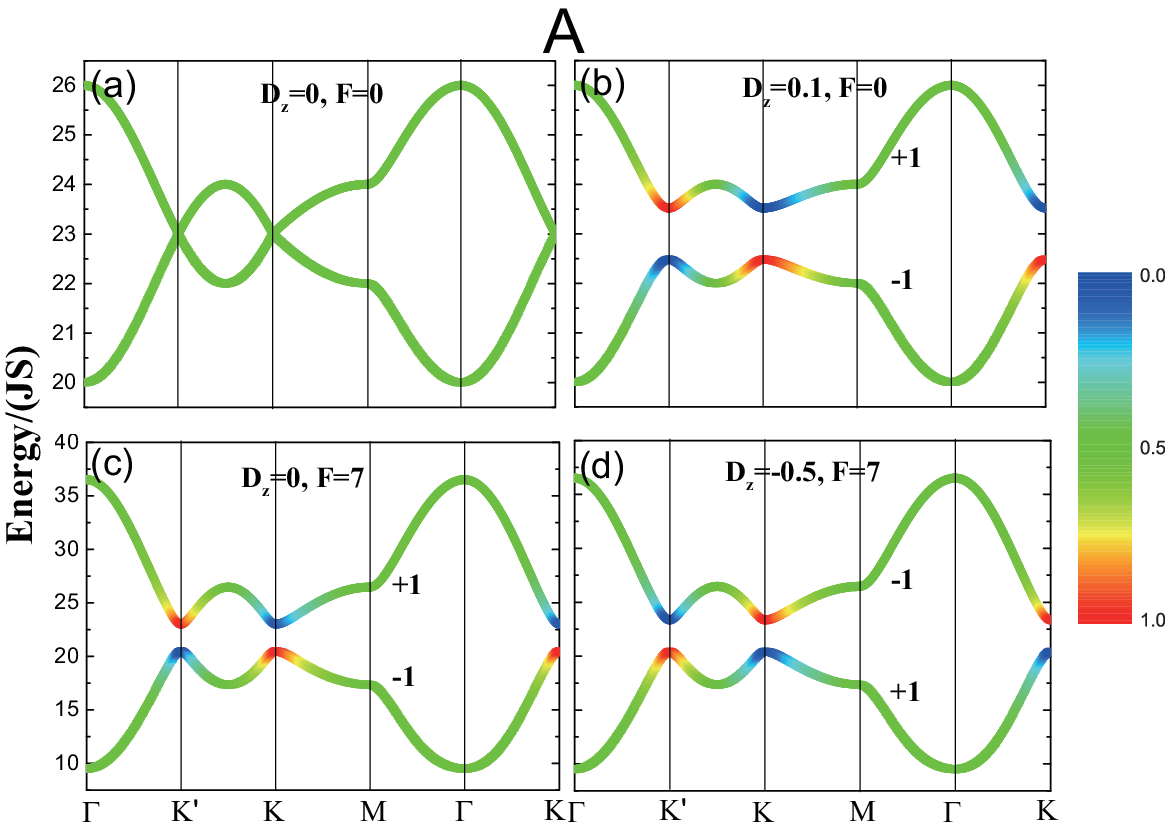}
\caption{(Color online) Pseudo-orbital (sublattice A, {\it i.e.} $\alpha$-orbital) projection at different parameters: (a) $D_{z}$=0, $F$=0; (b) $D_{z}$=0.1, $F$=0; (c) $D_{z}$=0, $F$=7; (d) $D_{z}$=$-$0.5, $F$=7. The red, blue and green colors denote $\alpha$, $\beta$ and proportionally mixed orbital occupations, respectively. The symbols $+/-$1 denote the Chern number of the corresponding bands.}
\label{Fig4}
\end{figure}

In general, the phase transition from linear Dirac state to topological insulating state is accompanied by the band inversion due to the gap opening. Thus we will analyze the band character, since there exists a gap evolution process within the topological phase transition. As is known, there are two sublattices A and B on the honeycomb lattice \cite{EPL103-47010}, which can be regarded as two pseudo-orbitals $\alpha$ and $\beta$, respectively. The orbital-resolved band dispersions at different cases for both trivial and non-trivial phases are compared, as displayed in Fig.~\ref{Fig4}. For the Dirac phase without anisotropic interactions ($D_{z}$=0, $F$=0), it is a proportionally mixed orbital feature in the whole BZ. While for the nontrivial phase owning to the DMI or PDI, the conduction and valence bands possess the opposite orbital feature at the energy gap of $K$ and $K^{'}$ points. Meanwhile, comparing $K$ with $K^{'}$, they have opposite orbital occupations, that is, opposite chiralities.

Moreover, the TP-I and TP-II phases indeed undergo band (orbital character) inversion corresponding to the topological phase transition, similar to that in topological insulators. Notice that this is an analogy to the valley effect in the electronic system, indicating the magnon valley\cite{PRB102-075407,NJP23-053022} degree of freedom can be achieved through manipulating the anisotropic interactions. In fact, the DMI and PDI can be precisely regulated by strain, hydrostatic pressure, electric and magnetic fields, {\it etc}. As a consequence, this may open a research field of valleymagnonics just like valleytronics \cite{NRM1-16055} in the electronic systems. We expect that ferrimagnetism can be used to realize a valley manipulation by lifting the pseudo-orbital degeneracy, which breaks the sublattice symmetry of A and B sites in a honeycomb lattice.

\subsection{\it Spin structure factor}
The obtained band dispersions can be compared with the experiments.
As is known that, neutron scattering is often used to provide magnon spectrum experimentally, and its cross section is proportional to spin structure factor $S^{\mathrm{in}}(k, \omega)$. The formula of spin structure factor \cite{PRL97-017003} is given by
\begin{equation}
\begin{aligned}
S^{\mathrm{in}}(\mathbf{k}, \omega)=& S \sum_{\lambda}\left|\left\langle\lambda\left|\sum_{m=\{1,2,3,4\}}\xi_{\mathbf{k} m} \alpha_{m}^{\dagger}\right|0\right\rangle\right|^{2}\times \delta\left(\omega-\omega_{f}\right) .
\end{aligned}
\end{equation}
where $\lambda$ is the band index, $|0\rangle$ is the vacuum state and $|\lambda\rangle$ is the final state, $\xi_{\lambda m}$is the $m$-th component of eigenvector $\alpha_{m}^{\dagger}|0\rangle$.

\begin{figure}[htbp]
\hspace*{-2mm}
\centering
\includegraphics[trim = 0mm 0mm 0mm 0mm, clip=true, angle=0, width=1.0 \columnwidth]{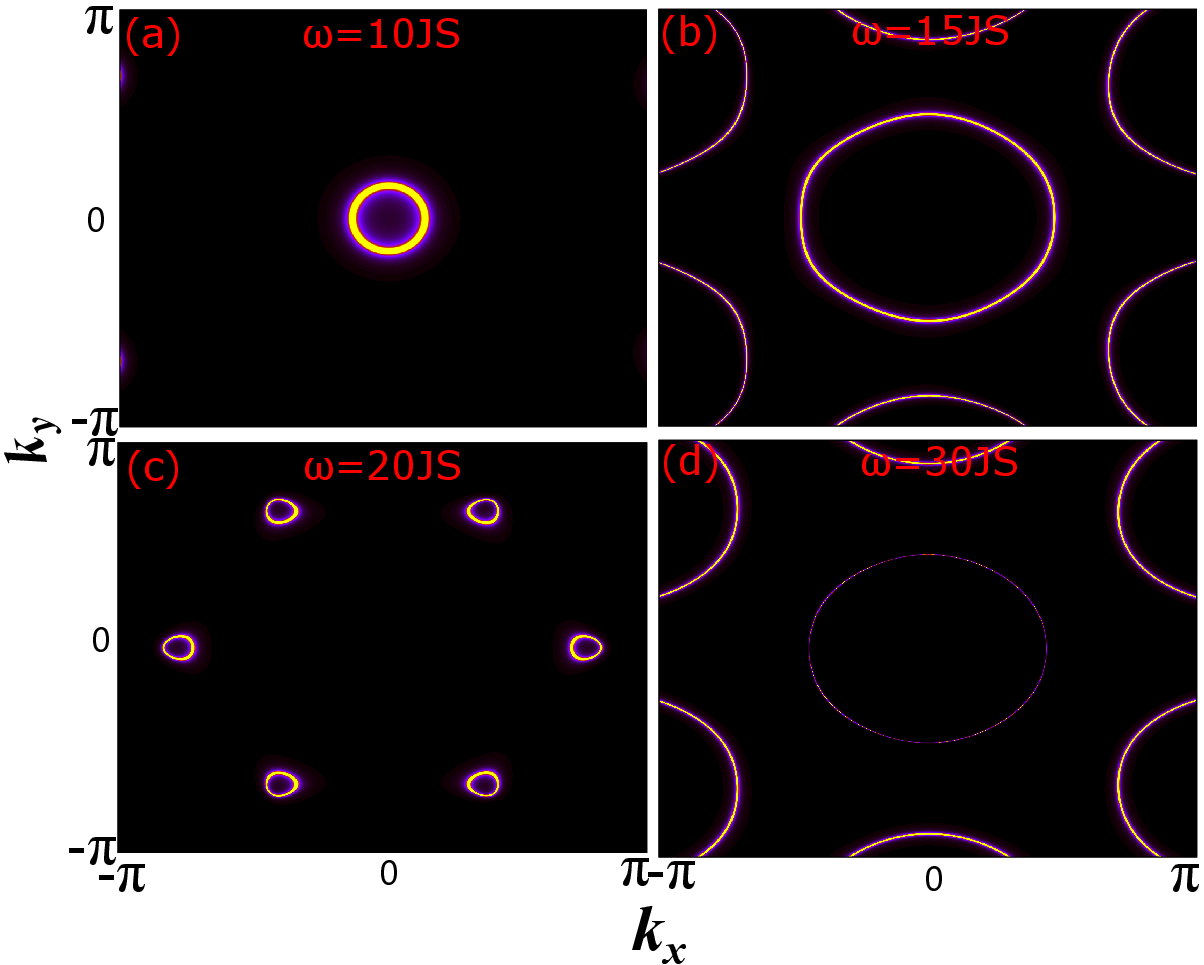}
\caption{(Color online) Spin structure factor with parameters  $D_{z}$=$-$0.5 and $F$=7 for different cut frequencies: (a) ${\omega}$=10$JS$, (b) ${\omega}$=15$JS$, (c) ${\omega}$=20$JS$, (d) ${\omega}$=30$JS$.}
\label{Fig5}
\end{figure}

Here we only show the result at parameters with $D_{z}$=$-$0.5, $F$=7, as shown in Fig.~\ref{Fig5}. Considering that the spin structure factor is a function of frequency, so several cut frequencies $\omega=10JS$, 15$JS$, 20$JS$ and 30$JS$ are plotted, which can be compared with the further experiments.

\subsection{\it Berry curvature and Chern number}
\begin{figure}[htbp]
\hspace*{-2mm}
\centering
\includegraphics[trim = 0mm 0mm 0mm 0mm, clip=true, angle=0, width=1.0 \columnwidth]{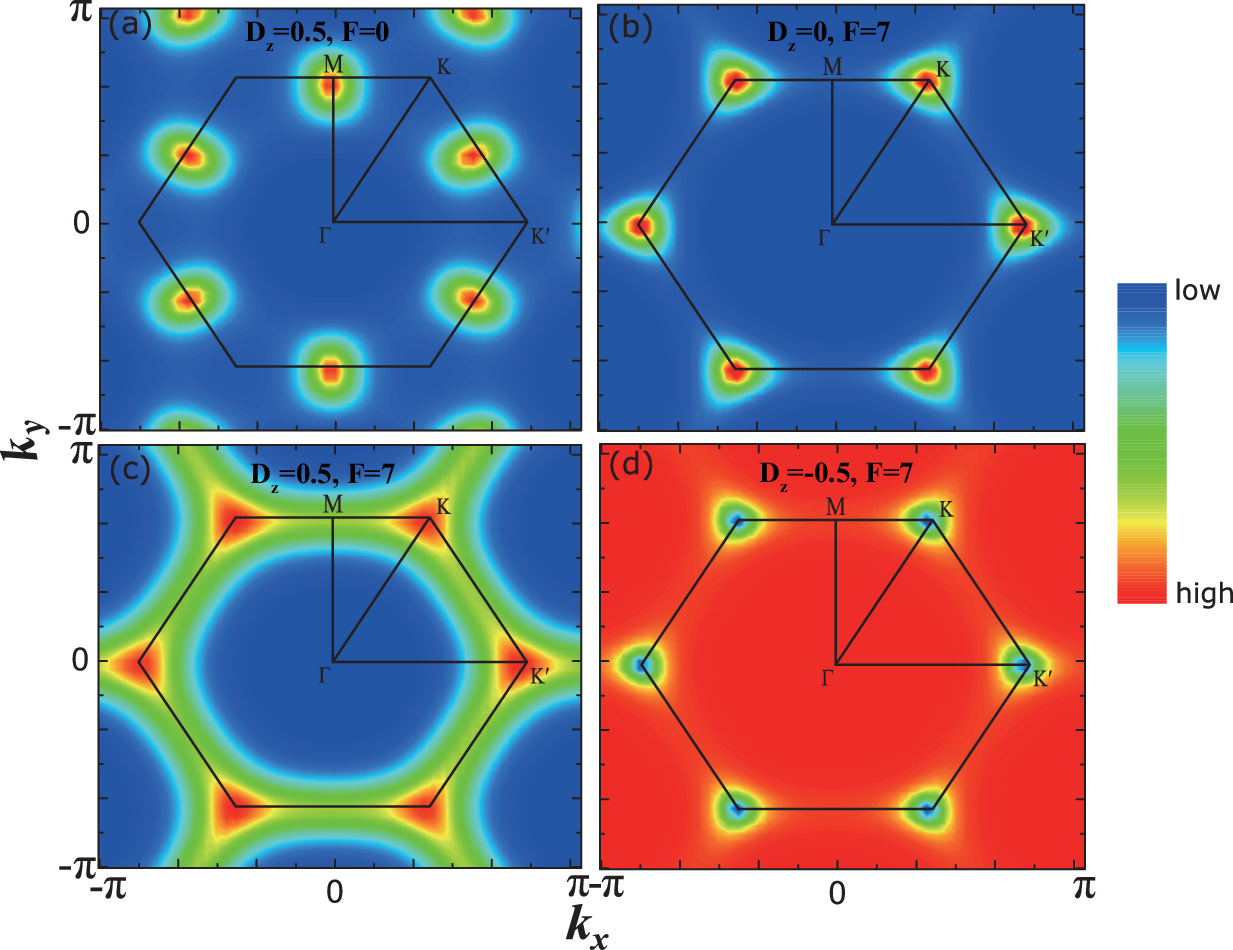}
\caption{(Color online) Berry curvature of valence (lower) band at different parameters: (a) $D_{z}$=0.5, $F$=0; (b) $D_{z}$=0, $F$=7; (c) $D_{z}$=0.5, $F$=7; (d) $D_{z}$=$-$0.5, $F$=7.}
\label{Fig6}
\end{figure}

The properties of the topological magnons are associated with the Berry curvature and topological invariants, such as Chern number, thus we investigate them in detail, and distinguish the roles of the anisotropic DMI and PDI. The momentum $k$-space Berry curvature \cite{PRSLA392-45} of the $n$-th magnon band is defined as \cite{PRB87-174427}
\begin{equation}\label{2}
B_n(\mathbf{k})=i\epsilon_{\mu \nu} \operatorname{Tr}\left[(1-Q_n)\left(\partial_{k_\mu} Q_n\right)\left(\partial_{k_\nu} Q_n\right)\right],
\end{equation}
which is different from the fermionic system due to the bosonic diagonalization. $Q_{n}$ is a projection operator in the vector space $Q_n=T_{\mathbf{k}} \Gamma_n \sigma_3 T_{\mathbf{k}}^{\dagger} \sigma_3$. $T_{\mathbf{k}}$ is a paraunitary matrix satisfied $T_{\mathbf{k}}^{\dagger} H_{\mathbf{k}} T_{\mathbf{k}}=\left[\begin{array}{ll}
E_{\mathbf{k}} & \\
& E_{-\mathbf{k}}
\end{array}\right]$ and $\Gamma_n$ is a diagonal matrix with +1 for every diagonal component and zero otherwise. Because the sum of total Berry curvature over the whole BZ is zero, the  conduction (upper) and valence (lower) bands just have different signs, thus only the Berry curvature of the valence band is shown in Fig.~\ref{Fig6} at different parameters for comparison. Obviously, though both DMI and PDI can induce magnon topological insulators, their Berry curvatures are distinctly different. The high-intensity regions of Berry curvature contributed from the pure DMI (PDI) mainly locate at $M$ point and center between $K$ and $K^{'}$ points (at $K$ and $K^{'}$ points) in the BZ. Due to the different $k$-space distributions from DMI and PDI, in the presence of both of them, the Berry curvature has a mixing character. In addition, the TP-II topological phase possesses opposite intensity distribution of Berry curvature compared with the TP-I phase.

On the other hand, Chern number, one of the topological invariants, is an integral of the Berry phase of a band over the first BZ, which can be computed by
\begin{equation}\label{2}
C_{n}=\frac{1}{2\pi}\int_{BZ}d\mathbf{k}B_{n}(\mathbf{k})
\end{equation}
As displayed in Figs.~\ref{Fig2} and \ref{Fig3}, the conduction and valence bands has opposite Chern numbers. It is worth noting that, the topological phase transition corresponds to the sign reversion of the Chern numbers between TP-I and TP-II phases.

\subsection{\it Edge state}
The topologically protected edge state is one of the important consequences of topological states. The edge states are calculated by constructing a slab model. The slab is set to be finite along the $y$ direction, but infinite along the $x$ direction, {\it i.e.} zigzag edge case, as shown in Fig.~\ref{Fig1}. The Hamiltonian of the principal layer is a 4$\times$4 matrix, thus when $N$ slabs are chosen, a $4N\times4N$ matrix with the basis $\Psi_{\mathbf{k}}^{\dag}=(a_{1,\mathbf{k}}^{\dag}, a_{1,-\mathbf{k}}, b_{1,\mathbf{k}}^{\dag}, b_{1,-\mathbf{k}}, ..., a_{N,\mathbf{k}}^{\dag}, a_{N,-\mathbf{k}}, b_{N,\mathbf{k}}^{\dag}, b_{N,-\mathbf{k}})$ are constructed.
\begin{equation}
H_{edge}=\left(\begin{array}{ccccc}
H_{11} & H_{12} & 0 & 0 & 0 \\
H_{21} & H_{22} & H_{23} & 0 & 0 \\
0 & H_{32} & H_{33} & \ddots & 0 \\
0 & 0 & \ddots & \ddots & H_{N-1,N} \\
0 & 0 & 0 & H_{N,N-1} & H_{NN}
\end{array}\right)
\end{equation}
where $H_{11}$ is the Hamiltonian of the first intra-slab, $H_{12}$ includes the interactions between the first and second slabs, {\it etc}.
\begin{figure}[htbp]
\hspace*{-2mm}
\centering
\includegraphics[trim = 0mm 0mm 0mm 0mm, clip=true, angle=0, width=1.0 \columnwidth]{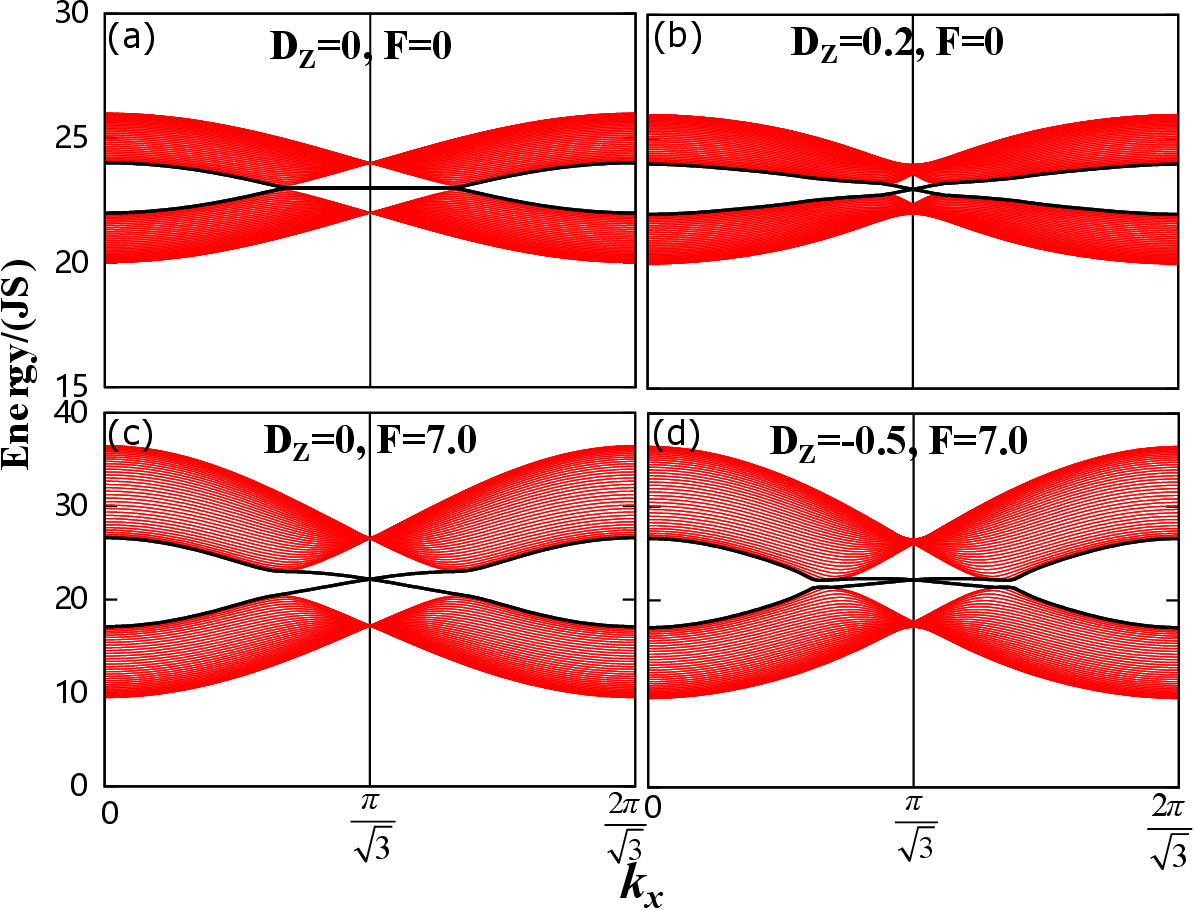}
\caption{(Color online) Edge states at zigzag edge for different anisotropic interaction parameters: (a) $D_{z}$=0, $F$=0; (b) $D_{z}$=0.2, $F$=0; (c) $F$=7.0, $D_{z}$=0; (c) $D_{z}$=$-$0.5, $F$=7. The edge state is degenerate for topologically trivial state, and linearly crossed for topologically non-trivial state.}
\label{Fig7}
\end{figure}
From Fig.~\ref{Fig7}, it can be found that in the absence of DMI and PDI, the edge states are trivial and degenerated flat bands. In the presence of DMI and PDI, the topologically non-trivial edge states with Dirac dispersion within the gap are induced as that in topological insulators of electronic systems.

\subsection{\it Magnon thermal Hall conductivity}
\begin{figure}[htbp]
\hspace*{-2mm}
\centering
\includegraphics[trim = 0mm 0mm 0mm 0mm, clip=true, angle=0, width=1.0 \columnwidth]{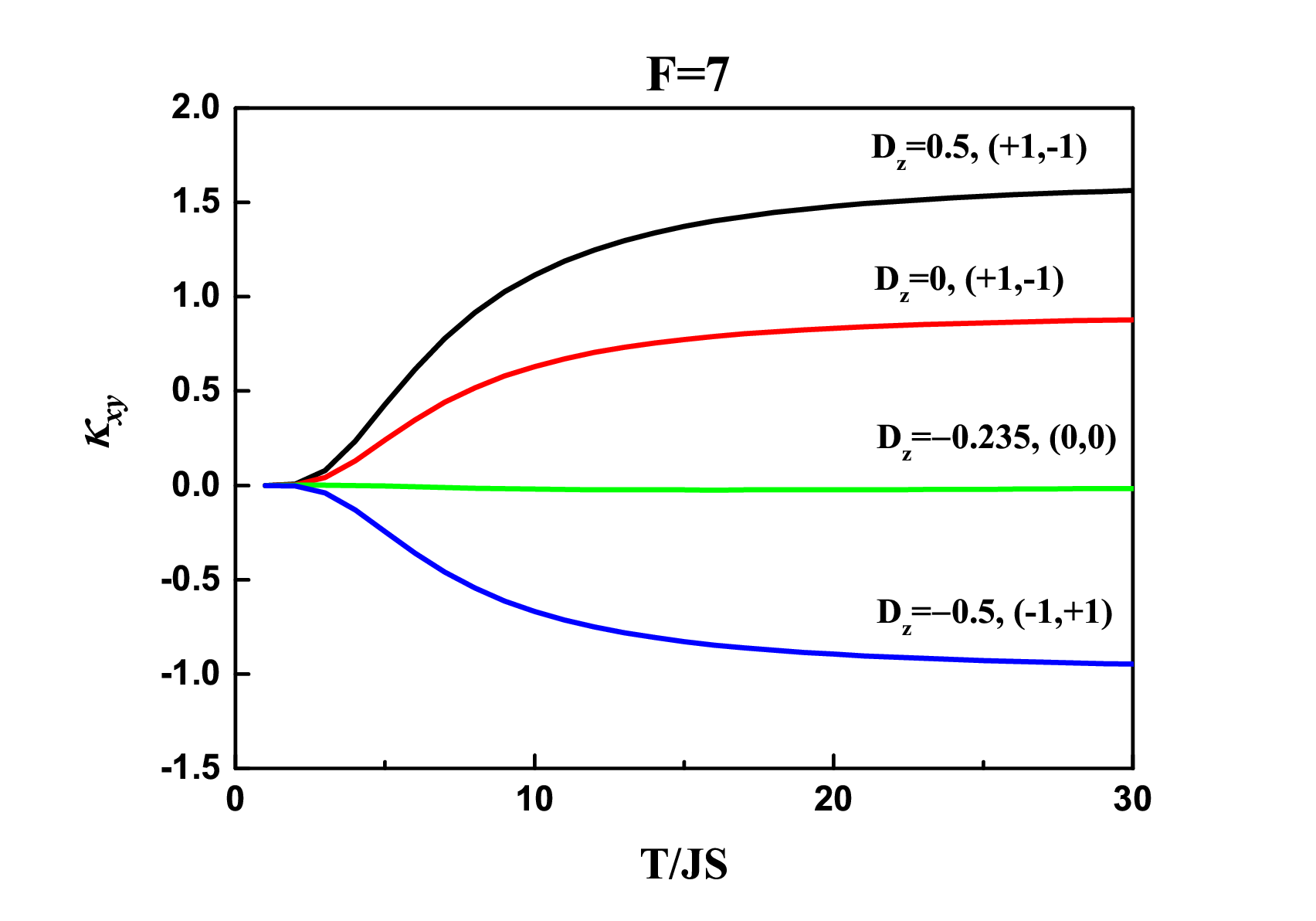}
\caption{(Color online) Magnon thermal Hall conductivity at different anisotropic interaction parameters. For simplicity, $F=7$ is fixed, only $D_{z}$ is varied. The symbols ($+/-$1, $-/+$1) denote the Chern number of conduction and valence bands, respectively.}
\label{Fig8}
\end{figure}

As a consequence of the topological magnons, the magnon thermal Hall effect can be directly observed experimentally. Here the magnon thermal Hall conductivities are calculated for comparing the different magnon topological phases. The magnon thermal Hall conductivity of the $z$ component can be written as \cite{PRB89-054420}
\begin{equation}
\kappa^{xy}=-\frac{k^{2}_{B}T}{(2\pi)^{2}}\sum\limits_{\lambda}\int_{BZ}(c_{2}(n_{\lambda})-\frac{\pi^{2}}{3})B_{n}(k)d\mathbf{k},
\end{equation}
where $n_{\lambda}=(e^{E_{\lambda}(k)/{k_{B}T}}-1)^{-1}$ is the Bose distribution function, and $c_{2}(x)=(1+x)(ln\frac{1+x}{x})^{2}-(lnx)^{2}-2Li_{2}(-x)$, in which $Li_{2}(-x)$ is the polylogarithm function. As shown in Fig.~\ref{Fig8}, it is obviously noticed that the signs of the magnon thermal Hall conductivities of the TP-I and TP-II phases are opposite, consistent with the topological phase transition. However, the thermal Hall conductivity of the critical intermediate phase between TP-I and TP-II phases is almost zero at $D_{z}=-0.235$ for $F=7$, indicating a topologically trivial phase rather than a non-trivial one. Accordingly, based on the sign change of the thermal Hall conductivity, one can realize the new magnonics devices with quantum manipulation.

\section{Conclusion}
In summary, the effect of the interplay of the anisotropic DMI and PDI on the properties of topological magnons in a ferromagnetic honeycomb lattice is investigated. Although the DMI and PDI can individually induce topological magnon insulator, the interplay of them, which behaves as either cooperating or competing, also introduces new topological phase as well as a chirality-dependent topological phase transition accompanied with a sign change of the topological invariants, {\it i.e.} chirality-reversal. Especially, the emergence of topological magnon phase transitions in the coexistence of DMI and PDI does not necessarily depend on large anisotropic parameters, but on the ratio of them. Therefore, this novel topological phase transition can be spontaneously realized in realistic materials. Indeed, different anisotropic magnetic interactions have their unique impact on topological magnons. And here we found that the anisotropic DMI and PDI play distinctly different roles in topological phase transition. The corresponding magnon thermal Hall conductivity undergoes a significant change, suggesting an obviously experimental evidence for identification. Consequently, from the topological phase diagram dependence on the anisotropic exchange interactions, two types of magnon topological phases in ferromagnetic honeycomb can be distinguished. On the other hand, this chirality-reversal magnon topological phase from the competitive anisotropic interactions is possibly hidden in the experiments due to the complicated effects. In addition, the band inversion originating from the pseudo-orbital reversal provides a magnon valley degree of freedom analogous to an electronic system, which is a good platform for realizing the topological magnonics devices. One can expect that applying strain on a ferromagnetic honeycomb lattice can not only precisely adjust the relative strength between DMI and PDI, but also lift the degeneracy of the valley at $K$ and $K'$ through breaking the lattice symmetry, hence realize the magnon valley regulation. The magnon valley degree of freedom deserves to be further investigated.
All the results may be identified by the neutron scattering and thermal transport experiments on magnons. Further exploration of the manipulation of the topological magnons deserves to be performed in realistic correlated magnetic materials and to be extended to the other lattices and magnetic interactions.

\acknowledgements
We acknowledge support from the National Natural Science Foundation of China under Grant Nos. 11974354, 11774350, and 11574315. Numerical calculations are performed at the Center for Computational Science of CASHIPS.

\end{document}